\documentclass[12pt]{iopart}

\usepackage{graphicx}

\begin{document}

\title{Strange Exotic States and Compact Stars}

\author{Irina Sagert, Mirjam Wietoska, 
J\"urgen Schaffner--Bielich\footnote{invited talk given at the SQM2006
  conference, UCLA, March 26--31, 2006}}

\address{Institut f\"ur Theoretische Physik/Astrophysik,
J. W. Goethe Universit\"at,
D--60054~Frankfurt am Main, Germany}

\ead{schaffner@astro.uni-frankfurt.de}

\begin{abstract} 
  We discuss the possible appearance of strange exotic multi-quark
  states in the interior of neutron stars and signals for the
  existence of strange quark matter in the core of compact stars. We
  show how the in-medium properties of possible pentaquark states are
  constrained by pulsar mass measurements. The possibility of
  generating the observed large pulsar kick velocities by asymmetric
  emission of neutrinos from strange quark matter in magnetic fields
  is outlined.
\end{abstract}

\section{Introduction}

In the spirit of this session on exotic quark states, strange quark
matter and astrophysics, we are going to discuss the interrelation of
these seemingly different topics. First, we give a brief discussion on
multiquark states, in detail we address the properties of multiquark
states with strangeness and charm. We consider light pentaquarks with
strangeness then heavy pentaquarks and multi-quark state with charm.
We summarise the current understanding of the in-medium properties of
pentaquarks in cold dense matter and extrapolate it to the case of
neutron star matter in beta-equilibrium. The possible appearance of
various pentaquark states and its implications for the global
properties of neutron stars is analysed and confronted with recent
pulsar mass measurements.

In the second part of this talk, we present newly suggested signals for
the possible presence of strange quark matter in the core of compact
stars. In particular, we focus on the phenomenon of pulsar kicks, which
has received considerable attention recently. The recent detailed
measurements of velocities of nearby pulsars in our galaxy clearly
shows, that speeds in excess of 1000 km/s are involved, much higher than
the typical velocity of ordinary nearby stars. So far a generally
accepted explanation for this effect is not at hand. In addition, we
briefly comment on recent developments on the relation of gamma-ray
bursts with strange quark matter and the recent gravitational wave
signals derived for the presence of strange quark matter in compact
stars.


\section{Multiquark States and Neutron Stars}

\subsection{Multi-Quark States: Some History (incomplete)}
  
The possible existence of multi-quark states, i.e.\ hadronic states
with more than three quarks, has been a topic of investigations for
decades since it was mentioned as a possibility within the quark model
by Gell-Mann in 1964 \cite{Gell64}. The first detailed predictions of
a bound multi-quark state was put forward by Jaffe in 1977: four-quark
states with strangeness (qs\=q\=s) \cite{Jaffe77a}, and the famous
H-dibaryon, a six-quark state with equal numbers of up-, down-, and
strange quarks (uuddss) \cite{Jaffe77}. Shortly afterwards, the
possibility of having bound multi-quark systems with heavy quarks
(QQ\=q\=q) were considered, too, in particular for the heavy
tetraquark systems, which consist of two light antiquarks and two
heavy quarks (or vice versa) \cite{Ader:1981db}. It was shown that
heavy tetraquarks must be bound in the limit of infinite heavy quarks.
Interestingly, pentaquark states were first proposed with a charm
anti-quark (qqqs\=c) in 1987 \cite{Lipkin87,Gig87}. Lipkin considered
symmetry arguments for the four light quarks within SU(6) to show that
the pentaquark can be a bound state, i.e.\ the decay to a baryon and a
meson is energetically forbidden. These arguments can not be applied
to a pentaquark state with an antistrange quark and four light quarks
consisting of up- and down-quarks only (qqqq\=s). Exactly such a
pentaquark state was proposed as a sharp resonant state by Diakonov,
Petrov, Polyakov in 1997 \cite{Diakonov:1997mm} on the basis of the
chiral soliton model. Key entry for its stability is the hypercharge,
not the spin-colour structure of the interaction. The light
pentaquarks are grouped within an antidecuplet, the lightest
candidate, the $\Theta^+$ being on the top of the triangle. The light
pentaquarks can be also described by the diquark model of Jaffe and
Wilczek \cite{Jaffe:2003sg}, which takes into account the recent
developments of colour superconductivity and strong diquark pairing in
QCD.

As is well known in this community, the light pentaquark $\Theta^+$ has
received enormous interest as several experimental groups reported a
signal for observing a pentaquark state starting in 2003. The mass is
1.54 GeV and it has a small width of less than 1 MeV. The experimental
situation is conflicting today, the experimental status of confidence as
given by the particle data group went down to only two stars in 2005.
But the experimental situation is not hopeless and we refer to the
corresponding overview talk given by Nakano at this meeting for more
details. We note that the antiparticle partner, the $\bar\Theta^-$ has
been observed by the ZEUS collaboration \cite{Chekanov:2004kn}.

In addition, other multiquark states have been reported recently also
for heavy-ion collisions. A possible signal for the $\Theta^{++}$ and
its antiparticle $\bar\Theta^{--}$ has been seen by the STAR
collaboration in d-Au collisions (but not in p-p or Au-Au collisions)
\cite{Salur:2004kz,Kabana:2005tp} and for the H-dibaryon
\cite{Vernet2005}. NA49 reports on detecting a signal for the doubly
strange $\Xi(1860)$ pentaquark (ssdd\=u), which is a member of the light
pentaquark antidecuplet \cite{Alt:2003vb}. Finally, a charmed pentaquark
$\Theta_c(3100)$ (uudd\=c) state was detected by the H1 collaboration
\cite{Ozerov:2005yz}.

\subsection{Multi-quark states with charm: charmlets}

Charm quarks have a stabilising effect for multiply charmed quark
states. First, due to its heavy mass a shallower potential is needed
to bind the system. Second, the colour-electric interaction between
heavy quarks is highly attractive. Third, the light quarks can combine
to a favourable antisymmetric flavour state for non-colour singlet
states. The latter fact is for example the reason why the charmed
pentaquark gets some additional binding energy compared to its
possible decay products \cite{Lipkin87}, while the first two apply for
e.g.\ the heavy tetra-quark states \cite{Ader:1981db}.

Bulk charmed strange quark matter turns out to be deeply bound within
the MIT bag model when including effects from single-gluon exchange
\cite{SV97} even for large values of the MIT bag constant. Therefore,
charmed strange quark matter can form bound objects even for the case
when ordinary strange quark matter (without charm quarks) is unbound.
For the description of multiply charmed quark states, as e.g.\ for the
charmonium states, it is essential to include explicitly the
colour-electric interaction terms. This modified bag model can
describe reasonably well both, light nonstrange and strange hadrons as
well as charmed hadrons with a slightly larger bag constant
\cite{SV97} compared to the standard MIT bag model \cite{DJJK75}. A
shell-model calculation of multiply charmed and strange quark states,
however, does not give bound candidates. The primary reason for this
result originates from the effectively added Casimir energy which
favours ordinary hadronic states over many-quark states. Possible
candidates close to being bound are the hexaquark states
\{cssuud\}$^+$, \{cssudd\}$^0$, \{ccsuud\}$^{++}$, \{ccsuud\}$^+$, and
\{ccssud\}$^+$. Those exotic objects can be produced in collisions of
heavy-ions at relativistic bombarding energies. The production rates
at RHIC amount to $10^{-3}$ to $10^{-2}$ per event within a simple
coalescence model \cite{SV97}.  Direct charm measurements will be
possible with the PHENIX and STAR detectors at RHIC with their upgrade
in the near future using e.g.\ micro-vertex detectors! A unique signal
would be a double star event on typical timescales of the weak decay
of charmed hadrons, i.e.\ within 100 micron or so. Note, that the same
event pattern, a double star, lead to the discovery of hypernuclei in
1953 \cite{Danysz53}!

\subsection{Pentaquarks in the medium: $\Theta^+$ hyponuclei?}

Hypernuclei are bound systems of nucleons with usually one or two
hyperons, the more general term strange hadronic matter applies to bound
systems of nucleons and multiple hyperons in general (see e.g.\
\cite{SG00} and references therein). In particular $\Lambda$ hypernuclei
have been studied extensively experimentally and theoretically and
attractive optical potentials of $U(\Lambda)\approx 30$ MeV have been
extracted from hypernuclear data. It is straightforward to imagine, that
the light strange pentaquarks can form similarly bound objects with
nuclei. So called hyponuclei (see e.g.\ the historical note in
\cite{Gal:2004cx}) have been considered in the literature for $\Theta^+$
pentaquarks. The simple quark model gives a highly attractive potential
for the $\Theta^+$, in fact the real part of the optical potential can
be below the critical value of $U_c(\Theta^+)=-105$ MeV so that the
$\Theta^+$ can not decay to a $\Lambda$ and a kaon anymore and is
stabilised in nuclei \cite{Miller:2004rj}! Within the relativistic
mean-field model, potential depths for the $\Theta^+$ of $U(\Theta^+) =
-37.5$ to $-90$ MeV at normal nuclear matter density $n_0$ have been
considered \cite{Zhong:2004hn,Zhong:2005dt}. QCD sum rule estimates
arrive at similar values of $U(\Theta^+) = -40$ to $-90$ MeV
\cite{Navarra:2004bj}. A more general hadronic SU(3) approach allows for
$U(\Theta^+) = -60 $ to $-120$ MeV if including coupling to two-meson
channels \cite{Cabrera:2004yg}.  Finally, a quark-mean field model
favours $U(\Theta^+) \approx -50$ MeV \cite{Shen:2004pa}. Interestingly,
the presence of the $\Theta^+$ in reactions of kaons with nuclei can
explain the long-standing puzzle of its missing reactivity which was put
forward by Gal and Friedman \cite{Gal:2004cx} and outlined in detail
within a coupled-channel calculation by Tolos, Cabrera, Ramos, and Polls
\cite{Tolos:2005jg}.

\subsection{Pentaquarks in Neutron Stars}

Now it is well known, that exotic and strange particles can be present
in high-density matter in the core of neutron stars.  Hyperons for
example appear around twice normal nuclear matter density (see e.g.\ 
the discussion in \cite{Scha02} and references given therein). While
the maximum mass of neutron stars is larger than $2M_\odot$ for matter
consisting of nucleons and leptons only, it is reduced considerably
below that value when the effect of hyperonisation is taken into
account \cite{GM91}. The maximum mass can be further reduced when more
exotic species are present at high densities. Every new degree of
freedom reduces in principle the pressure for a given energy density
for a free Fermi gas. Strong interactions for quarks, however, can
invalidate that argument and can lead to compact stars with strange
quark matter cores with maximum masses close to $2M_\odot$
\cite{Alford:2006vz}.

In the following we explore the possibility that pentaquarks can
affect the maximum mass of compact stars and derive limits for their
maximum attainable attractive potential in dense matter for the case
of purely hadronic compact stars (see \cite{Wietoska2005} for
details). In fact, we find that the presence of $\Theta^+$ leads to
an overall reduction of the maximum mass of compact stars, similar to
the case of hyperons, so that such constraints can be derived. We
consider the relativistic mean-field model with nucleons, leptons and
hyperons, where the interaction is mediated by the exchange of scalar
($\sigma$, $\sigma^*$) and vector mesons ($\omega$, $\rho$, $\phi$)
using SU(6) symmetry relations for the coupling constants of the
vector mesons and adjusting the potential depth with the scalar
coupling constants appropriately \cite{Scha93,Scha94,SM96,SG00}. The
coupling constants for pentaquarks are chosen similarly. We use the
naive quark counting rule for fixing the relative vector coupling
strength. The scalar coupling constant is determined by choosing a
potential depth of the pentaquark at normal nuclear matter density and
varied accordingly. We adopt a negative vector coupling strength as
advocated in \cite{Navarra:2004bj} using QCD sum rules. Note, that in
any case the pentaquark, like the other baryons, feels an overall
repulsive (vector) potential at high densities, which garanties the
stability of matter (otherwise matter will continously increase its
density and collapse to a black hole)!

The model under study predicts that hyperons are present in neutron
star matter around twice normal nuclear matter density \cite{SM96}.
Assuming a potential depth of the $\Theta^+$ pentaquark at normal
nuclear matter density of $U(\Theta^+)=-100$ MeV, it will appear
around 4$n_0$ in beta-equilibrated dense matter. For the maximum mass
configuration, the fraction of pentaquarks in the core of the compact
star amounts to about five percent. 

The $\Theta^+$ can appear even for a repulsive potential of
$U(\Theta^+)=+50$ MeV at $n_0$. The reason is that the repulsive
potential felt by the pentaquark rises less steeply with density
compared to the nucleon one, so that it will become comparably smaller
which allows for pentaquarks to emerge in sufficiently high density
matter. The presence of pentaquarks reduces the overall pressure, so
that the maximum mass decreases. The change is, however, quite small,
so that the constraints on the pentaquark in-medium potential are
moderate. Recent pulsar mass measurements of pulsar binaries with
white dwarfs point towards $M>1.6 M_\odot$, see the presentation given
by Ingrid Stairs at this meeting.  For a pulsar mass of $M>1.6
M_\odot$, the potential depth of the pentaquark must be $U(\Theta^+) >
-190$ MeV (for parameter set GM1 of \cite{GM91}) which is a rather
weak constraint.

The same line of reasoning can be adopted for other possible
pentaquark states. In particular, we expect that negatively charged
states are favoured in compact star matter as charge balance with the
positively charged protons can be more easily accomplished. Indeed,
pulsar mass constraints for a hypothetical negatively charged
pentaquark $\Theta^-$ in matter lead to much stronger limits:
$U(\Theta^-) > -55 \dots -125$ MeV for $M>1.44 M_\odot$ and
$U(\Theta^-) > 0 \dots -85$ MeV for $M>1.6 M_\odot$. The range
indicates the values found for various different parameter sets, see
\cite{Wietoska2005}. The composition of compact stars with $\Theta^-$
pentaquarks shows even more drastic changes compared to an ordinary
neutron star. The negatively charged $\Theta^-$ appears already at
$2n_0$ (for $U(\Theta^-) = -100$ MeV and set GM1) being the first
exotic and strange component in beta-equilibrated matter! The maximum
density attainable for stable compact star configurations is reduced
to only $3.3n_0$. The doubly negative charged pentaquark $\Xi^{--}$ is
a member of the pentaquark antidecuplet. However, the effect of having
twice the negative charge is compensated by the proposed (and
measured) much heavier mass of the $\Xi^{--}$ of around 1860 MeV.
Constraints on its in-medium potential turn out to be rather weak,
only hypothetical small masses and large attractive potentials are
incompatible with pulsar mass measurements.


\section{Signals for Strange Quark Matter in Compact Stars}

Now we turn to the second topic of this presentation, discussing
recently suggested signals for the presence of strange quark matter in
compact stars. We will be rather brief without any intention to give
an overview on this topic (which is impossible for this
booming research field) and focus on topics which are heavily biased
towards personal work. 

In any case, hunting down strange quark matter in the heavens is
certainly coming of age! Some recently suggested signals include besides
'exotic' mass-radius relation of compact stars and the enhanced cooling
of neutron stars, gamma-ray bursts by the transition to strange quark
matter, gravitational wave signals of the collisions of compact stars
with quark matter, and pulsar kicks by asymmetric emission of neutrinos
from quarks. We will concentrate on the latter issue in the following
with some small remarks on the other two.  

\subsection{Pulsar kicks and strange quark matter}

There are nearly three-hundred measurements of pulsar velocities
recently and it has been observed that the pulsars move out of the
centre of the supernova remnant with high space velocities of up to
1600 km/s (see \cite{Hobbs:2005yx} and references therein). The
highest directly measured kick velocity has been determined to be
$1080\pm 100$ km/s \cite{Chatterjee:2005mj}. So pulsars are speeding
through our Galaxy! 

We note in passing that the distribution of pulsar kick velocities has
been found in some earlier work to be bimodal \cite{Arzoumanian2002}
which has been connected to the coexistence of neutron stars and quark
stars by Bombaci and Popov \cite{Bombaci:2004nu}, where the quark stars
experience a second kick due to the phase transition from hadronic to
quark matter (see also \cite{Bombaci:2005nu}).

For three pulsars (Crab, Vela and pulsar B0656+14) the observed kick
axis is closely aligned with the rotational axis.  Several explanations
for the puzzling pulsar kick phenomenon have been put forward as (with
some comments in brackets):

\begin{itemize}
\item the asymmetric emission of sterile neutrinos from a deformed
  neutrino-sphere due to magnetic fields \cite{Kusenko:1996sr} (this
  relies on the properties of hypothetical exotic particles)
\item parity-violating neutrino emission in strong magnetic fields in
  nuclear matter \cite{Horowitz:1997fb} (only surface emission is
  possible, large magnetic fields at the surface are needed)
\item spin--1 colour superconductivity produces asymmetric emission of
  neutrinos \cite{Schmitt:2005ee} (operates only below a temperature of 1
  MeV, so that the resulting kick velocities are too small)
\item electromagnetic vortices in colour--superconducting quark matter
  \cite{Berdermann:2005yn} let neutrinos propagates preferable along the
  magnetic field axis (however, there are no electromagnetic vortices
  in the standard colour-superconducting phases considered)
\item anisotropy of the hydrodynamic shock wave in neutrino-driven
  supernova explosion \cite{Scheck:2003rw} (which needs an artificially
  increased $\nu$--interaction for successful simulation of an
  explosion)
\end{itemize}

Pulsars are born in core-collapse supernovae, first as hot
proto-neutron stars. The energy stored in neutrinos amounts to 99
percent of the total energy released in a supernova (of type II). The
momentum necessary for the observed pulsar kick velocities is about
one percent of the one stored in neutrinos, so that an asymmetry of
one percent for neutrino emission is sufficient for a successful kick!
However, besides the energetics (enough kick?) also the anisotropy
(enough polarisation?) and the efficiency of the neutrino rocket (do
they come out?) has to be checked carefully. In the following, we
address these issues for getting a kick out of strange quark matter by
asymmetric neutrino emission in strong magnetic fields (for details
see \cite{Sagert2006}).

Electron capture on quarks is parity-violating, so that in principle
anisotropic production of neutrinos for degenerate matter can occur.
If all neutrinos released from the proto-neutron star are emitted
along one direction, an initial temperature of just about $T=5$ MeV is
needed for a sufficient pulsar kick of 1000 km/s which follows from
momentum conservation and the total thermal energy available.  Hence,
the kick originates from the early proto-neutron star evolution, say
within the first minute of the supernova event, and not from later
stages.

For asymmetric neutrino emission, we consider here the case of
polarisation by ultrastrong magnetic fields. The measured surface
magnetic fields of pulsars are typically around $B\simeq 10^{12}$
Gauss. In certain cases, for so called magnetars, surface magnetic fields
of up to $10^{15}$ Gauss have been inferred. Moreover, the magnetic
field in the quark core can be much larger than at the surface so that
magnetic fields in quark matter of up to $10^{18}$ Gauss are possible.
The electrons are moving then in Landau levels, even in degenerate
matter. The lowest Landau level is fully polarised, electron capture
on these electrons produce neutrinos which are emitted along one
specific direction.

The last criterion for a successful neutrino rocket involves the
mean-free paths of neutrinos in dense and hot matter (do they come
out?). The scattering reaction $\nu + n \to \nu + n$ and the capture
reaction $\nu_e + n \to e^- + p$ occur in neutron matter and lead to
mean-free paths of just a few meters for a temperature of $T=5$ MeV
\cite{Reddy:2002xc}. For unpaired quark matter, similar processes give a
mean-free path of around 100 meters. Processes with fully paired
quarks, as present in the colour-flavour-locked phase (CFL), are
exponentially suppressed, so that CFL quark matter should be
transparent to neutrinos. However, the CFL phase violates baryon
number conservation and breaks a $U(1)$ symmetry. The corresponding
massless excitation, the $H$, reduces the mean-free path back to the
small values for ordinary quark matter, unfortunately
\cite{Reddy:2002xc}!

So now the problem is that neutrinos have to get out without
rescattering to preserve the anisotropy, but the mean-free path is too
short, shorter than the typical radius of a neutron star of 10 km at a
temperature of $T=5$ MeV. When effects from the H are ignored, the
mean-free path increases exponentially with the gap proportional to
$\exp (-\Delta/T)$. But the thermal energy (the specific heat) stored
decreases then exponentially, so that not enough momentum is available
for a sufficiently high kick velocity! What is needed then is some
mechanism to increase the specific heat of colour superconducting
quark matter, maybe by including effects from Goldstone modes.

\subsection{Gamma-ray bursters and strange quark matter}

The phenomenon of gamma-ray bursts are one of the most energetic
events observed in the sky, the energy released is similar to the ones
of supernovae. About one gamma-ray burst is measured per day. The most
promising sources considered so far involve colliding neutron stars,
stars collapsing to black holes or collapsing compact stars!
Recently, a signal from quiescent gamma-ray bursters has been proposed
by Pagliara and Drago \cite{Drago:2005rc}. They extracted a subsample
with long quiescent periods (more than 40 seconds) from the BATSE
catalog.  Interestingly, the bursts before and after quiescence
exhibit a similar characteristic which points to a dormant inner
engine rather than to physics related to e.g.\ the pulsar wind. The
properties of long quiescent gamma-ray burster hint then at a collapse
of a neutron star with two phase transitions: the first transition
proceeds from nuclear matter to quark matter, the second one to
colour--superconducting matter releasing the gap energy of gapped
strange quark matter!

\subsection{Gravitational wave signal from strange quark matter?}

There has been a dramatic advancement in recent years in the field of
simulating and modelling the gravitational wave signal from compact
stars which takes into account effects from the presence of quark
matter. Binary neutron star mergers have been simulated including a
quark matter equation of state at high densities \cite{Oechslin:2004yj}.
The different equations of state affect clearly the Fourier spectrum
of the gravitational wave emitted. Binary strange star collisions have
been studied in \cite{Limousin:2004vc}. Compared to ordinary neutron
stars, strange stars cause higher frequencies in the gravitational
wave signal before 'touch-down' due to being more compact in the
cases studied by the authors. A collapse of a normal neutron star to a
compact star with a quark matter core has been modelled in
\cite{Lin:2005zd}. Again, the gravitational wave pattern exhibits
sensitivities to the underlying equation of state!


\section{Summary}

We have discussed multiquark states, with strangeness and with charm,
and used pulsar mass measurements to derive constraints on the
in-medium properties of light pentaquark states. Multiply charmed
multiquark states have unusual properties and can be bound by strong
colour-electric forces. Double prong decay patterns can signal their
existence, if bound, and can be measured with the future tracking
devices for charmed particles of the PHENIX and STAR detectors at
RHIC. One should not miss this oportunity and look for it! Light
pentaquarks, i.e.\ five-quark states with strangeness, can be present
in the core of neutron stars and reduce the maximum mass. Hence, their
in-medium properties can be constrained by pulsar mass measurements.
Present limits give only loose bounds, which become much tighter with
only a slightly increased pulsar mass limit. We are looking forward
to ongoing and future pulsar scans and continued observations of
binary pulsars which will reduce the uncertainty considerably in the
near future!

The presence of strange quark matter in compact stars can signal
itself in various astrophysical observables. The puzzling observation
of high velocity pulsars might be related to the presence of strange
quark matter in pulsars. The conditions for a successful neutrino
rocket, the asymmetric emission of neutrinos due to polarised
electrons in magnetic fields, have been examined in detail.  Large
magnetic fields are needed, which might be present in the quark core
of compact stars. Enough kick is reached if the asymmetric emission of
neutrinos starts at high enough temperature which is only reached in
the early evolution of the proto-neutron star stage, i.e.\ within a
minute after the supernova collapse. At these temperatures, however,
the neutrino mean-free path is shorter than the radius of the compact
star, in particular in neutron matter. Sufficiently large values are
reached for colour-superconducting quark matter in the CFL phase only,
if effects from a massless excitation, the so called H, are ignored.
However, then the specific heat and the available thermal energy is
reduced exponentially so that not enough kick can be produced. Hence,
our present understanding of high-density QCD and
colour-superconductivity does not give large enough mean-free paths or
sufficiently large heat capacities to allow for a successful neutrino
rocket. More detailed investigations are needed to pursue this issue
in more detail.

The burst pattern of gamma-ray bursters with long quiescence times
might indicate the presence of two separate phase transitions in
compact stars: one from nuclear matter to normal quark matter then the
second one to colour-superconducting quark matter. Also gravitational
wave patterns can exhibit the presence of strange quark matter in
compact stars either from compact star mergers or from collapsing
neutron stars.

Last but not least much more astrophysical data is on the horizon poised for
discovery. There are lots of opportunities for strange quark matter
physics. Certainly, the cross-talk with observers will be crucial and
vital for a successful advancement of strange astrophysics!


\ack
This work is supported in part by the Gesellschaft f\"ur
Schwerionenforschung mbH, Darmstadt, Germany.

\section*{References}

\bibliographystyle{revtex}
\bibliography{all,literat,sqm2006}

\begin{thebibliography}{10}
\providecommand*{\bibinfo}[2]{#2}
\providecommand*{\eprint}[1]{#1}
\providecommand*{\url}[1]{#1}
\bibitem{Gell64}
\bibinfo{author}{M.~Gell-Mann}, \bibinfo{journal}{Phys. Rev. Lett.}
  \bibinfo{volume}{\textbf{12}}, \bibinfo{pages}{155} (\bibinfo{date}{1964}).
\bibitem{Jaffe77a}
\bibinfo{author}{R.~L. Jaffe}, \bibinfo{journal}{Phys. Rev. D}
  \bibinfo{volume}{\textbf{15}}, \bibinfo{pages}{267} (\bibinfo{date}{1977}).
\bibitem{Jaffe77}
\bibinfo{author}{R.~L. Jaffe}, \bibinfo{journal}{Phys. Rev. Lett.}
  \bibinfo{volume}{\textbf{38}}, \bibinfo{pages}{195} (\bibinfo{date}{1977}),
  erratum: ibid 38 (1977) 617.
\bibitem{Ader:1981db}
\bibinfo{author}{J.~P. Ader}, \bibinfo{author}{J.~M. Richard}, and
  \bibinfo{author}{P.~Taxil}, \bibinfo{journal}{Phys. Rev.}
  \bibinfo{volume}{\textbf{D25}}, \bibinfo{pages}{2370} (\bibinfo{date}{1982}).
\bibitem{Lipkin87}
\bibinfo{author}{H.~L. Lipkin}, \bibinfo{journal}{Phys. Lett. B}
  \bibinfo{volume}{\textbf{195}}, \bibinfo{pages}{484} (\bibinfo{date}{1987}).
\bibitem{Gig87}
\bibinfo{author}{C.~Gignoux}, \bibinfo{author}{B.~Silvestre-Brac}, and
  \bibinfo{author}{J.~M. Richard}, \bibinfo{journal}{Phys. Lett. B}
  \bibinfo{volume}{\textbf{193}}, \bibinfo{pages}{323} (\bibinfo{date}{1987}).
\bibitem{Diakonov:1997mm}
\bibinfo{author}{D.~Diakonov}, \bibinfo{author}{V.~Petrov}, and
  \bibinfo{author}{M.~V. Polyakov}, \bibinfo{journal}{Z. Phys.}
  \bibinfo{volume}{\textbf{A359}}, \bibinfo{pages}{305} (\bibinfo{date}{1997}),
  \eprint{hep-ph/9703373}.
\bibitem{Jaffe:2003sg}
\bibinfo{author}{R.~L. Jaffe} and \bibinfo{author}{F.~Wilczek},
  \bibinfo{journal}{Phys. Rev. Lett.} \bibinfo{volume}{\textbf{91}},
  \bibinfo{pages}{232003} (\bibinfo{date}{2003}), \eprint{hep-ph/0307341}.
\bibitem{Chekanov:2004kn}
\bibinfo{author}{S.~Chekanov} \emph{et~al.} (\bibinfo{collaboration}{ZEUS}),
  \bibinfo{journal}{Phys. Lett.} \bibinfo{volume}{\textbf{B591}},
  \bibinfo{pages}{7} (\bibinfo{date}{2004}), \eprint{hep-ex/0403051}.
\bibitem{Salur:2004kz}
\bibinfo{author}{S.~Salur} (\bibinfo{collaboration}{STAR}),
  \bibinfo{journal}{nucl-ex/0403009}  (\bibinfo{date}{2004}).
\bibitem{Kabana:2005tp}
\bibinfo{author}{S.~Kabana}, \bibinfo{journal}{J. Phys.}
  \bibinfo{volume}{\textbf{G31}}, \bibinfo{pages}{S1155}
  (\bibinfo{date}{2005}), \eprint{hep-ex/0503019}.
\bibitem{Vernet2005}
\bibinfo{author}{R.~Vernet}, Ph.D. thesis, Universit\'e Louis Pasteur,
  Strasbourg, France (\bibinfo{date}{2005}).
\bibitem{Alt:2003vb}
\bibinfo{author}{C.~Alt} \emph{et~al.} (\bibinfo{collaboration}{NA49}),
  \bibinfo{journal}{Phys. Rev. Lett.} \bibinfo{volume}{\textbf{92}},
  \bibinfo{pages}{042003} (\bibinfo{date}{2004}), \eprint{hep-ex/0310014}.
\bibitem{Ozerov:2005yz}
\bibinfo{author}{D.~Ozerov} (\bibinfo{collaboration}{H1 and ZEUS}),
  \bibinfo{journal}{PoS} \bibinfo{volume}{\textbf{HEP2005}},
  \bibinfo{pages}{088} (\bibinfo{date}{2006}).
\bibitem{SV97}
\bibinfo{author}{J.~Schaffner-Bielich} and \bibinfo{author}{A.~P. Vischer},
  \bibinfo{journal}{Phys. Rev. D} \bibinfo{volume}{\textbf{57}},
  \bibinfo{pages}{4142} (\bibinfo{date}{1997}).
\bibitem{DJJK75}
\bibinfo{author}{T.~DeGrand}, \bibinfo{author}{R.~L. Jaffe},
  \bibinfo{author}{K.~Johnson}, and \bibinfo{author}{J.~Kiskis},
  \bibinfo{journal}{Phys. Rev. D} \bibinfo{volume}{\textbf{12}},
  \bibinfo{pages}{2060} (\bibinfo{date}{1975}).
\bibitem{Danysz53}
\bibinfo{author}{M.~Danysz} and \bibinfo{author}{J.~Pniewski},
  \bibinfo{journal}{Phil. Mag.} \bibinfo{volume}{\textbf{44}},
  \bibinfo{pages}{348} (\bibinfo{date}{1953}).
\bibitem{SG00}
\bibinfo{author}{J.~Schaffner-Bielich} and \bibinfo{author}{A.~Gal},
  \bibinfo{journal}{Phys. Rev. C} \bibinfo{volume}{\textbf{62}},
  \bibinfo{pages}{034311} (\bibinfo{date}{2000}).
\bibitem{Gal:2004cx}
\bibinfo{author}{A.~Gal} and \bibinfo{author}{E.~Friedman},
  \bibinfo{journal}{Phys. Rev. Lett.} \bibinfo{volume}{\textbf{94}},
  \bibinfo{pages}{072301} (\bibinfo{date}{2005}), \eprint{nucl-th/0411052}.
\bibitem{Miller:2004rj}
\bibinfo{author}{G.~A. Miller}, \bibinfo{journal}{Phys. Rev.}
  \bibinfo{volume}{\textbf{C70}}, \bibinfo{pages}{022202}
  (\bibinfo{date}{2004}), \eprint{nucl-th/0402099}.
\bibitem{Zhong:2004hn}
\bibinfo{author}{X.~H. Zhong}, \bibinfo{author}{Y.~H. Tan},
  \bibinfo{author}{L.~Li}, and \bibinfo{author}{P.~Z. Ning},
  \bibinfo{journal}{Phys. Rev.} \bibinfo{volume}{\textbf{C71}},
  \bibinfo{pages}{015206} (\bibinfo{date}{2005}), \eprint{nucl-th/0408046}.
\bibitem{Zhong:2005dt}
\bibinfo{author}{X.~H. Zhong}, \bibinfo{author}{G.~X. Peng}, and
  \bibinfo{author}{P.~Z. Ning}, \bibinfo{journal}{Phys. Rev.}
  \bibinfo{volume}{\textbf{C72}}, \bibinfo{pages}{065212}
  (\bibinfo{date}{2005}), \eprint{/}.
\bibitem{Navarra:2004bj}
\bibinfo{author}{F.~S. Navarra}, \bibinfo{author}{M.~Nielsen}, and
  \bibinfo{author}{K.~Tsushima}, \bibinfo{journal}{Phys. Lett.}
  \bibinfo{volume}{\textbf{B606}}, \bibinfo{pages}{335} (\bibinfo{date}{2005}),
  \eprint{nucl-th/0408072}.
\bibitem{Cabrera:2004yg}
\bibinfo{author}{D.~Cabrera}, \bibinfo{author}{Q.~B. Li},
  \bibinfo{author}{V.~K. Magas}, \bibinfo{author}{E.~Oset}, and
  \bibinfo{author}{M.~J. Vicente~Vacas}, \bibinfo{journal}{Phys. Lett.}
  \bibinfo{volume}{\textbf{B608}}, \bibinfo{pages}{231} (\bibinfo{date}{2005}),
  \eprint{nucl-th/0407007}.
\bibitem{Shen:2004pa}
\bibinfo{author}{H.~Shen} and \bibinfo{author}{H.~Toki},
  \bibinfo{journal}{Phys. Rev.} \bibinfo{volume}{\textbf{C71}},
  \bibinfo{pages}{065208} (\bibinfo{date}{2005}), \eprint{nucl-th/0410072}.
\bibitem{Tolos:2005jg}
\bibinfo{author}{L.~Tolos}, \bibinfo{author}{D.~Cabrera},
  \bibinfo{author}{A.~Ramos}, and \bibinfo{author}{A.~Polls},
  \bibinfo{journal}{Phys. Lett.} \bibinfo{volume}{\textbf{B632}},
  \bibinfo{pages}{219} (\bibinfo{date}{2006}), \eprint{hep-ph/0503009}.
\bibitem{Scha02}
\bibinfo{author}{J.~Schaffner-Bielich}, \bibinfo{author}{M.~Hanauske},
  \bibinfo{author}{H.~St{\"o}cker}, and \bibinfo{author}{W.~Greiner},
  \bibinfo{journal}{Phys. Rev. Lett.} \bibinfo{volume}{\textbf{89}},
  \bibinfo{pages}{171101} (\bibinfo{date}{2002}), \eprint{astro-ph/0005490}.
\bibitem{GM91}
\bibinfo{author}{N.~K. Glendenning} and \bibinfo{author}{S.~A. Moszkowski},
  \bibinfo{journal}{Phys. Rev. Lett.} \bibinfo{volume}{\textbf{67}},
  \bibinfo{pages}{2414} (\bibinfo{date}{1991}).
\bibitem{Alford:2006vz}
\bibinfo{author}{M.~Alford}, \bibinfo{author}{D.~Blaschke},
  \bibinfo{author}{A.~Drago}, \bibinfo{author}{T.~Kl{\"a}hn},
  \bibinfo{author}{G.~Pagliara}, and \bibinfo{author}{J.~Schaffner-Bielich},
  \bibinfo{journal}{astro-ph/0606524}  (\bibinfo{date}{2006}).
\bibitem{Wietoska2005}
Mirjam Wietoska, Diploma Thesis, Goethe University, Frankfurt, Germany (2005);
  M.~Wietoska and J.~Schaffner-Bielich, manuscript in preparation.
\bibitem{Scha93}
\bibinfo{author}{J.~Schaffner}, \bibinfo{author}{C.~B. Dover},
  \bibinfo{author}{A.~Gal}, \bibinfo{author}{C.~Greiner}, and
  \bibinfo{author}{H.~St\"ocker}, \bibinfo{journal}{Phys. Rev. Lett.}
  \bibinfo{volume}{\textbf{71}}, \bibinfo{pages}{1328} (\bibinfo{date}{1993}).
\bibitem{Scha94}
\bibinfo{author}{J.~Schaffner}, \bibinfo{author}{C.~B. Dover},
  \bibinfo{author}{A.~Gal}, \bibinfo{author}{D.~J. Millener},
  \bibinfo{author}{C.~Greiner}, and \bibinfo{author}{H.~St\"ocker},
  \bibinfo{journal}{Ann. Phys. (N.Y.)} \bibinfo{volume}{\textbf{235}},
  \bibinfo{pages}{35} (\bibinfo{date}{1994}).
\bibitem{SM96}
\bibinfo{author}{J.~Schaffner} and \bibinfo{author}{I.~N. Mishustin},
  \bibinfo{journal}{Phys. Rev. C} \bibinfo{volume}{\textbf{53}},
  \bibinfo{pages}{1416} (\bibinfo{date}{1996}).
\bibitem{Hobbs:2005yx}
\bibinfo{author}{G.~Hobbs}, \bibinfo{author}{D.~R. Lorimer},
  \bibinfo{author}{A.~G. Lyne}, and \bibinfo{author}{M.~Kramer},
  \bibinfo{journal}{Mon. Not. Roy. Astron. Soc.}
  \bibinfo{volume}{\textbf{360}}, \bibinfo{pages}{963} (\bibinfo{date}{2005}),
  \eprint{astro-ph/0504584}.
\bibitem{Chatterjee:2005mj}
\bibinfo{author}{S.~Chatterjee} \emph{et~al.}, \bibinfo{journal}{Astrophys. J.}
  \bibinfo{volume}{\textbf{630}}, \bibinfo{pages}{L61} (\bibinfo{date}{2005}),
  \eprint{astro-ph/0509031}.
\bibitem{Arzoumanian2002}
\bibinfo{author}{Z.~{Arzoumanian}}, \bibinfo{author}{D.~F. {Chernoff}}, and
  \bibinfo{author}{J.~M. {Cordes}}, \bibinfo{journal}{Astrophys. J.}
  \bibinfo{volume}{\textbf{568}}, \bibinfo{pages}{289} (\bibinfo{date}{2002}),
  \eprint{astro-ph/0106159}.
\bibitem{Bombaci:2004nu}
\bibinfo{author}{I.~Bombaci} and \bibinfo{author}{S.~B. Popov},
  \bibinfo{journal}{Astron. Astrophys.} \bibinfo{volume}{\textbf{424}},
  \bibinfo{pages}{627} (\bibinfo{date}{2004}), \eprint{astro-ph/0405250}.
\bibitem{Bombaci:2005nu}
\bibinfo{author}{I.~Bombaci}, \bibinfo{journal}{J. Phys.}
  \bibinfo{volume}{\textbf{G31}}, \bibinfo{pages}{S825} (\bibinfo{date}{2005}).
\bibitem{Kusenko:1996sr}
\bibinfo{author}{A.~Kusenko} and \bibinfo{author}{G.~Segre},
  \bibinfo{journal}{Phys. Rev. Lett.} \bibinfo{volume}{\textbf{77}},
  \bibinfo{pages}{4872} (\bibinfo{date}{1996}), \eprint{hep-ph/9606428}.
\bibitem{Horowitz:1997fb}
\bibinfo{author}{C.~J. Horowitz} and \bibinfo{author}{G.~Li},
  \bibinfo{journal}{Phys. Rev. Lett.} \bibinfo{volume}{\textbf{80}},
  \bibinfo{pages}{3694} (\bibinfo{date}{1998}), \eprint{astro-ph/9705126}.
\bibitem{Schmitt:2005ee}
\bibinfo{author}{A.~Schmitt}, \bibinfo{author}{I.~A. Shovkovy}, and
  \bibinfo{author}{Q.~Wang}, \bibinfo{journal}{Phys. Rev. Lett.}
  \bibinfo{volume}{\textbf{94}}, \bibinfo{pages}{211101}
  (\bibinfo{date}{2005}), \eprint{hep-ph/0502166}.
\bibitem{Berdermann:2005yn}
\bibinfo{author}{J.~Berdermann}, \bibinfo{author}{D.~Blaschke},
  \bibinfo{author}{H.~Grigorian}, and \bibinfo{author}{D.~N. Voskresensky},
  \bibinfo{journal}{astro-ph/0512655}  (\bibinfo{date}{2005}).
\bibitem{Scheck:2003rw}
\bibinfo{author}{L.~Scheck}, \bibinfo{author}{T.~Plewa}, \bibinfo{author}{H.-T.
  Janka}, \bibinfo{author}{K.~Kifonidis}, and \bibinfo{author}{E.~M{\"u}ller},
  \bibinfo{journal}{Phys. Rev. Lett.} \bibinfo{volume}{\textbf{92}},
  \bibinfo{pages}{011103} (\bibinfo{date}{2004}), \eprint{astro-ph/0307352}.
\bibitem{Sagert2006}
Irina Sagert, Diploma Thesis, Goethe University, Frankfurt, Germany (2006);
  I.~Sagert and J.~Schaffner-Bielich, Proceedings of the 343rd WE-Heraeus
  Seminar on Neutron Stars and Pulsars, Bad Honnef, May 14-19, 2006, to be
  published as MPE report.
\bibitem{Reddy:2002xc}
\bibinfo{author}{S.~Reddy}, \bibinfo{author}{M.~Sadzikowski}, and
  \bibinfo{author}{M.~Tachibana}, \bibinfo{journal}{Nucl. Phys.}
  \bibinfo{volume}{\textbf{A714}}, \bibinfo{pages}{337} (\bibinfo{date}{2003}),
  \eprint{nucl-th/0203011}.
\bibitem{Drago:2005rc}
\bibinfo{author}{A.~Drago} and \bibinfo{author}{G.~Pagliara},
  \bibinfo{journal}{astro-ph/0512602}  (\bibinfo{date}{2005}).
\bibitem{Oechslin:2004yj}
\bibinfo{author}{R.~Oechslin}, \bibinfo{author}{K.~Uryu},
  \bibinfo{author}{G.~S. Poghosyan}, and \bibinfo{author}{F.~K. Thielemann},
  \bibinfo{journal}{Mon. Not. Roy. Astron. Soc.}
  \bibinfo{volume}{\textbf{349}}, \bibinfo{pages}{1469} (\bibinfo{date}{2004}),
  \eprint{astro-ph/0401083}.
\bibitem{Limousin:2004vc}
\bibinfo{author}{F.~Limousin}, \bibinfo{author}{D.~Gondek-Rosinska}, and
  \bibinfo{author}{E.~Gourgoulhon}, \bibinfo{journal}{Phys. Rev.}
  \bibinfo{volume}{\textbf{D71}}, \bibinfo{pages}{064012}
  (\bibinfo{date}{2005}), \eprint{gr-qc/0411127}.
\bibitem{Lin:2005zd}
\bibinfo{author}{L.-M. Lin}, \bibinfo{author}{K.~S. Cheng},
  \bibinfo{author}{M.~C. Chu}, and \bibinfo{author}{W.-M. Suen},
  \bibinfo{journal}{Astrophys. J.} \bibinfo{volume}{\textbf{639}},
  \bibinfo{pages}{382} (\bibinfo{date}{2006}), \eprint{astro-ph/0509447}.

\end{thebibliography}

\end{document}